\documentclass[review]{elsarticle}
\graphicspath{ {./figures/} }
\usepackage{hyperref}
\usepackage{float}
\usepackage{verbatim} 
\usepackage{apalike}
\restylefloat{figure}
\restylefloat{table}

\usepackage{booktabs}
\usepackage{multirow}

\journal{IEEE Access}

\biboptions{authoryear}

\begin{document}
\begin{frontmatter}

\title{MeWEHV: Mel and Wave Embeddings for Human Voice Tasks}

\author[label1,label2]{Andrés Carofilis \corref{cor1}}
\ead{andres.vasco@unileon.es}

\author[label2,label3]{Laura Fernández-Robles}
\ead{l.fernandez@unileon.es}

\author[label1,label2]{Enrique Alegre}
\ead{enrique.alegre@unileon.es}

\author[label1,label2]{Eduardo Fidalgo}
\ead{eduardo.fidalgo@unileon.es}

\cortext[cor1]{Corresponding author.}
\address[label1]{Dept. of Electrical, Systems and Automation Engineering. School of Industrial, Computer and Aerospace Engineering, Universidad de León, Campus Vegazana 24007, León, Spain}
\address[label2]{Researcher at INCIBE (Spanish National Cybersecurity Institute), León, Spain}
\address[label3]{Dept. of Mechanical, Computer and Aerospace Engineering, Universidad de León, Spain. }

\begin{abstract}
A recent trend in speech processing is the use of embeddings created through machine learning models trained on a specific task with large datasets. By leveraging the knowledge already acquired, these models can be reused in new tasks where the amount of available data is small. This paper proposes a pipeline to create a new model, called Mel and Wave Embeddings for Human Voice Tasks (MeWEHV), capable of generating robust embeddings for speech processing. MeWEHV combines the embeddings generated by a pre-trained raw audio waveform encoder model, and deep features extracted from Mel Frequency Cepstral Coefficients (MFCCs) using Convolutional Neural Networks (CNNs). 

We evaluate the performance of MeWEHV on three tasks: speaker, language, and accent identification. For the first one, we use the VoxCeleb1 dataset and present YouSpeakers204, a new and publicly available dataset for English speaker identification that contains 19607 audio clips from 204 persons speaking in six different accents, allowing other researchers to work with a very balanced dataset, and to create new models that are robust to multiple accents. For evaluating the language identification task, we use the VoxForge and Common Language datasets. Finally, for accent identification, we use the Latin American Spanish Corpora (LASC) and Common Voice datasets.

Our approach allows a significant increase in the performance of state-of-the-art models on all the tested datasets, with a low additional computational cost.

\end{abstract}

\begin{keyword}
XLSR-Wav2Vec2 \sep WavLM \sep HuBERT \sep YouSpeakers204 \sep Embeddings \sep Speech classification
\end{keyword}

\end{frontmatter}

\section{Introduction} \label{section:introduction}
\label{introduction}


Speech processing refers to analyzing human speech through voice audio signals. Some of the most important problems in this field, which are tackled in this paper, are language identification, accent identification, and speaker identification~\citep{DBLP:journals/asc/NassifSHNH21, DBLP:conf/icassp/HuangXYMQ21, DBLP:journals/corr/abs-2203-00328}.

On the one hand, language identification identifies the spoken language present in an audio file, and accent identification determines a person's region of origin based on the characteristic way and tone of the language used. We consider that the existence of very similar languages or accents, usually languages or accents with a common origin, poses a challenge for both tasks and requires the use of powerful machine learning models. In some speech processing tasks, there are useful datasets publicly available~\citep{garofolo1993timit, maclean_2018, ardila2020common, DBLP:conf/icassp/PanayotovCPK15, DBLP:conf/icassp/KahnRZKXMKLCFLS20, cui2013developing}, but the comparison of results on such datasets becomes difficult due to the lack of a common predefined experimental setup for training the models and the lack of previous research results to compare with, such as the case of accent detection in Spanish~\citep{guevara-2020-crowdsourcing}.

On the other hand, speaker identification consists of recognizing the identity of a person given an audio file with a person's voice. A problem in this field is that there is a lack of well-balanced English datasets, both in the number of audios per speaker and the number of speakers per accent~\citep{DBLP:conf/interspeech/NagraniCZ17, garofolo1993timit, cui2013developing, pratap2020mls, DBLP:conf/icassp/PanayotovCPK15, DBLP:conf/icassp/KahnRZKXMKLCFLS20}. This can lead to the creation of models that might not identify accents in real-world data effectively. The availability of a dataset with these features would allow the creation of more effective models on real-world problems and facilitate the integration and evaluation of multiple tasks, such as speaker identification and accent identification, simultaneously.

Most of the research focused on the three aforementioned tasks addresses them individually, and the proposed systems are usually evaluated for just one task~\citep{DBLP:journals/corr/abs-2202-12163, DBLP:journals/corr/abs-2203-00328, DBLP:journals/asc/NassifSHNH21, DBLP:journals/eswa/NassifSEVAP22,  DBLP:conf/icassp/HuangXYMQ21,DBLP:conf/interspeech/WangYHLX21}. However, some research addresses several of these problems based on the same pre-training model~\citep{shor20_interspeech, baevski2020wav2vec, DBLP:journals/taslp/HsuBTLSM21, DBLP:journals/corr/abs-2110-13900}.

A machine learning architecture capable of performing well on multiple speech processing tasks can use the knowledge acquired during training, with a large amount of data, and exploit it in new and diverse tasks. In this way, the model generated for a new task does not need to start from scratch, thus requiring less training data. By freezing the trained layers, fewer parameters would need to be trained, with a consequent reduction in the computational power required~\citep{shor20_interspeech, DBLP:journals/taslp/HsuBTLSM21, DBLP:journals/corr/abs-2110-13900}. 

Various techniques exist for reusing these models on specific tasks other than those for which they were initially trained. This field of research is known as \textit{transfer learning}~\citep{li2017learning}. Some models address the transfer learning problem by creating deep representations~\citep{baevski2020wav2vec, conneau21_interspeech, DBLP:journals/taslp/HsuBTLSM21, DBLP:journals/corr/abs-2110-13900}, also called \textit{embeddings}. An embedding represents a position in an abstract multidimensional space that encodes a meaningful internal representation of externally observed events. In these spaces, similar embeddings, or embeddings that have features in common, are close together, while less similar items are far apart~\citep{DBLP:conf/nips/KingmaMRW14}. The embeddings have been used in multiple domains, such as text, image, and speech processing, and can feed multiple systems for an individual task in each one~\citep{DBLP:journals/jair/Camacho-Collados18, DBLP:conf/emnlp/KielaB14}.

For the speech processing domain, there are models that address speech classification for one or more tasks using embeddings. For example, WavLM~\citep{DBLP:journals/corr/abs-2110-13900}, presented as a universal speech encoder, was tested in tasks such as speaker identification, and speech to text, among others. There are also models that, although they were developed for a specific task, are also capable of creating embeddings, so they can be reused in new tasks. This is the case of HuBERT~\citep{DBLP:journals/taslp/HsuBTLSM21} and Wav2Vec2~\citep{baevski2020wav2vec}, which are focused on English speech-to-text conversion, and XLSR-Wav2Vec2~\citep{conneau21_interspeech}, which is based on Wav2Vec2 but adds the possibility to work with multiple languages.


The embedding generation models mentioned above were trained with thousands of hours of audios recorded in multiple environments, resulting in models that can generate robust embeddings against background noise and different environmental conditions. This paper takes into account the advances achieved by this class of models and leverages them for speaker identification, accent identification, and language identification, by means of transfer learning. 

We develop and present a novel embedding enrichment procedure, which combines the outputs of two models. On the one hand, an embedding generation model from raw audios, which we refer to, in a general way, as \textit{wave encoder}. On the other hand, the outputs of a neural network (NN) fed by the Mel Frequency Cepstral Coefficients (MFCCs)~\citep{tiwari2010mfcc} of the raw audios, which have among its advantages the capability of error reduction and robustness to noise~\citep{kamarulafizam2007heart}. The main feature of MFCC is that it focuses on extracting relevant audio components to identify speech features, discarding, by filtering, other features such as background noise, pitch, loudness, and emotion, among others. Therefore, we call the NN an MFCC encoder.


The proposed architecture complements the high level of detail that the model exploits with the wave encoder, being this a non-imposed representation, and the extraction of relevant information through the MFCCs, as an imposed representation. The information contained in the raw audio may contain relevant information that may have been filtered out in the MFCC, and the MFCC provide the machine learning model with information on the most relevant parts of an audio, on which it should focus.

\begin{figure}[thpb]
  \centering
  {\includegraphics[width=\textwidth]{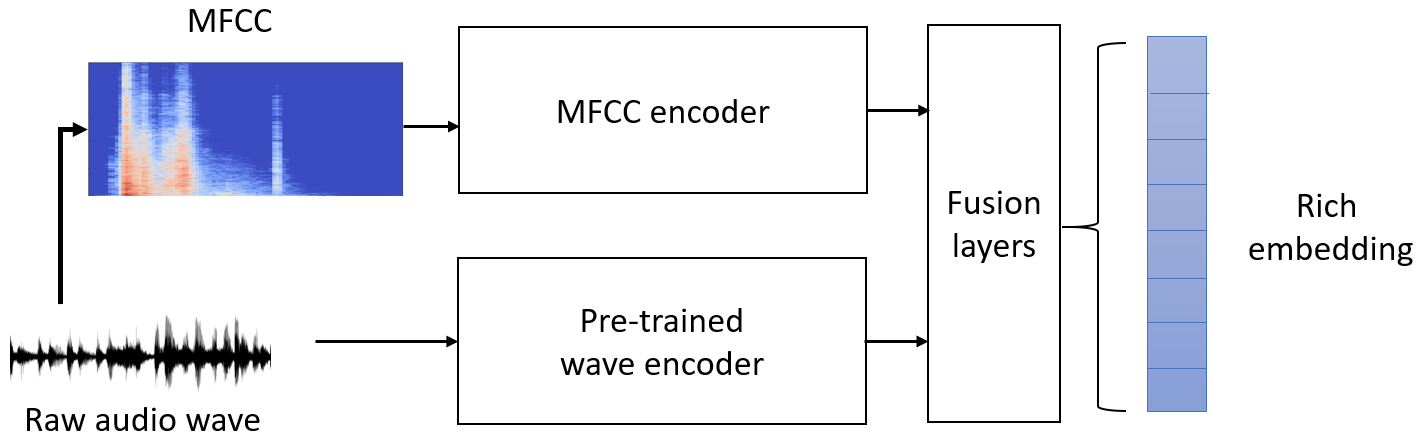}
  }
  \caption{A basic representation of the proposed architecture. It merges two types of representations and generates rich embeddings.}
  \label{figure:pipeline_model_basic}
\end{figure}

For the correct complementarity of both types of representations, we designed an architecture capable of interacting with them through a set of layers, including LSTM layers and attention mechanisms. In this way, we managed to overcome the results obtained by other state-of-the-art models, at the same time requiring only a small number of trainable parameters. Figure~\ref{figure:pipeline_model_basic} shows a basic scheme of the proposed architecture.

This paper provides the following main contributions:

\begin{itemize}
\item Proposal of a machine learning model architecture, which we call Mel and Wave Embeddings for Human Voice Tasks (MeWEHV), which is able to efficiently tackle multiple speech classification tasks, and achieves state-of-the-art results in all of them. Since our proposal reuses frozen weights from pre-trained models and requires a relatively low number of trainable parameters, our work can serve as a basis for other research with limited computational resources to retrain the huge models typically used in speech processing to generate new competitive models.
\item Proposal of a new pipeline for the generation of rich embeddings, by merging multiple representations, establishing a possible basis for new architectures that aim at improving large pre-trained models. 
\item Creation and presentation of a new speaker identification dataset, called YouSpeakers204, highly balanced in terms of speaker accents and gender, which was extracted from public YouTube videos, and used to create a speaker identification model. This dataset is made publicly available, along with the experimental setup to facilitate the comparison of results among different experiments\footnote{To get access to the YouSpeakers204 dataset, please contact us through our web page: https://gvis.unileon.es/contact-us}.
\item For the first time in the literature, we use the Latin American Spanish Corpora for the task of accent identification. We evaluated the MeWEHV architecture and other state-of-the-art baselines with this dataset, shared the experimental setup, and established baseline results for future research.
\item Application of this research to a real problem focused on the extraction of speaker information for the identification of offenders and victims. The research developed in this paper is part of the European Project GRACE, which seeks to apply machine learning techniques for the analysis and management of data for the fight against child sexual exploitation.
\end{itemize}


The remaining part of the paper is organized as follows: In Section~\ref{section:soa}, a review of the state of the art in the field of speech processing is presented. Then, in Section~\ref{section:youspeaker}, the information of the new YouSpeakers204 dataset is introduced. In Section~\ref{section:architecture}, the proposed architecture is described. In Section~\ref{section:experiments}, the rest of the datasets and the experimental setup are detailed. In Section~\ref{section:results}, the results obtained with each of the datasets are described. Finally, the discussion and the conclusions obtained are given in Sections~\ref{section:discussion} and~\ref{section:conclutions}, respectively.

\section{State of the art} \label{section:soa}

In the field of speech classification, multiple solutions have been developed for a single task. \cite{DBLP:journals/corr/abs-2202-12163} presented a language identification system based on conformer layers, and a temporal pooling mechanism, which was tested on their own dataset with 65 languages and achieved an accuracy up to 4.27\% higher than other approaches based on LSTM and transformers.

\cite{DBLP:journals/corr/abs-2203-00328} proposed BERT-LID, based on a conjunction network for phoneme recognition and BERT with a linear output layer. They evaluated their proposal on the datasets AP20-OLR~\citep{DBLP:conf/apsipa/LiZHLTWSY20}, TAL\_ASR, and a combination of the datasets THCHS-30~\citep{DBLP:journals/corr/WangZ15e} and TIMIT~\citep{garofolo1993timit}, achieving up to $5\%$ improvement in audios of more than three seconds and $18\%$ in audios of less than one second, with respect to models based on n-grams-SVM and x-vectors.

In speaker identification, \cite{DBLP:journals/asc/NassifSHNH21} introduced CASA-GMM-CNN model, in which they seek to clean a noisy audio through a Computational Auditory Scene Analysis (CASA), then make a classification of emotions through a GMM-CNN, and the output of both components feed another GMM-CNN in charge of identifying the speaker. They tested their approach on SUSAS~\citep{DBLP:conf/interspeech/HansenB97}, Arabic Emirati Speech Database~\citep{DBLP:journals/nca/ShahinNH20}, RAVDESS~\citep{livingstone2018ryerson}, and Fluent Speech Commands~\citep{DBLP:conf/interspeech/LugoschRITB19} datasets, achieving an improvement in accuracy of up to $59.37\%$ with respect to other state-of-the-art works.

\cite{DBLP:journals/eswa/NassifSEVAP22} presented another speaker identification model based on capsule networks, which is composed of two convolutional layers and one capsule layer, and it was compared using standard CNNs, random forests, GMM-DNNs, and SVMs as baseline models, on the Arabic Emirati Speech Database, SUSAS, and RAVDESS datasets. This model achieved improvements of up to $9.98\%$, $10.95\%$, and $9.81\%$ accuracy, respectively, with respect to the best baseline model.

In accent identification, \cite{DBLP:conf/icassp/HuangXYMQ21} presented AISpeech-SJTU, an accent identification system that is powered by Phone Posteriorgrams and data augmented by text-to-speech synthesis systems. They evaluated their proposal in the Interspeech-2020 Accented English Speech Recognition Challenge~\citep{DBLP:conf/icassp/ShiYLLFWQX21}, achieving an average accuracy of 83.63\%, the highest score of the challenge.

Transfer learning and domain transfer have been extensively studied in machine learning~\citep{DBLP:journals/pieee/ZhuangQDXZZXH21}. Recent research related to transfer learning in audio processing has mainly focused on methods for learning deep representations, also known as embeddings~\citep{shor20_interspeech}. These embeddings are generated to store relevant information of an audio wave, through its representation in a latent space, to be later used in the learning of a new specific task.

\cite{DBLP:conf/interspeech/WangYHLX21} presented an accent identification model generated from a pre-trained speech-to-text model, to which transfer learning was applied to be reused in their new task. To evaluate their proposal, they used the AP20-OLR dataset, achieving a reduction of up to $10.79\%$ in the EER compared to other approaches based on x-vectors and i-vectors.

One powerful model focused on the generation of embeddings is TRILL~\citep{shor20_interspeech}, which was trained with a subset of the AudioSet dataset~\citep{DBLP:conf/icassp/GemmekeEFJLMPR17}, and subsequently evaluated in different domains by applying transfer learning and fine-tuning. The results achieved with TRILL were, in most cases, superior to those of the state of the art, and in other cases, close to them, being able to highlight its performance in speaker identification, with an accuracy of 17.9\% on the VoxCeleb1 dataset~\citep{DBLP:conf/interspeech/NagraniCZ17}, 94.1\% for language identification on the VoxForge dataset (5.7\% improvement)~\citep{maclean_2018}, 91.2\% for command identification on the Speech Commands dataset~\citep{DBLP:journals/corr/abs-1804-03209} (0.1\% improvement), among others.

Other embedding generation models are the Wav2Vec2~\citep{baevski2020wav2vec} model, which focused on English speech-to-text conversion, and XLSR-Wav2Vec2~\citep{conneau21_interspeech} model. XLSR-Wav2Vec2 is based on Wav2Vec2 but has been adapted for speech-to-text conversion in 53 languages, where the use of embeddings is useful to adapt the model to the different languages. To train the XLSR-Wav2Vec2 model, the MLS~\citep{pratap2020mls}, CommonVoice~\citep{ardila2020common}, and BABEL~\citep{cui2013developing} datasets were used. The XLSR-Wav2Vec2 model is fed by the raw audio waves and was able to achieve a word error rate reduction of $72\%$ compared to other published results on the Common Voice dataset, and $16\%$ compared to the state-of-the-art results on BABEL.

\cite{DBLP:journals/taslp/HsuBTLSM21} presented a new self-supervised approach for embedding generation based on BERT, called HuBERT. HuBERT uses an offline clustering step to provide aligned target labels for a BERT-like prediction loss. The HuBERT model matches or improves the performance of Wav2Vec2 on Librispeech~\citep{DBLP:conf/icassp/PanayotovCPK15} and Libri-Light~\citep{DBLP:conf/icassp/KahnRZKXMKLCFLS20} datasets, achieving WER improvement of up to $19\%$.

\cite{DBLP:journals/corr/abs-2110-13900} presented the WavLM model extending the HuBERT framework for speech-to-text and denoising modeling, which enables pre-trained WavLM models to perform well on both speech-to-text and non-speech-to-text tasks. To achieve this, some WavLM inputs are noisy/overlapping speech simulations and the expected outputs are the original speech labels. In addition, they optimized the model structure and training data of HuBERT and Wav2Vec2. The model was tested in the SUPERB Challenge~\citep{DBLP:conf/interspeech/YangCCLLLLSCLHT21} achieving an overall score $3.16\%$ higher than HuBERT and $4.95\%$ higher than Wav2Vec2.

Apart from the models focused on speech processing, there are also models for general audio processing, such as the PANN model. The PANN model~\citep{DBLP:journals/taslp/KongCIWWP20} was trained on the AudioSet dataset and evaluated using transfer learning and fine-tuning, in general content audio classification tasks. For environmental sound classification and audio taggings, PANN yielded accuracies of 94.7\% and 96.0\% on the ESC-50~\citep{DBLP:conf/mm/Piczak15} and the MSoS~\citep{DBLP:conf/icassp/KroosBCHJDWCP19} datasets, respectively, surpassing the state-of-the-art results. 

For acoustic scene classification, PANN was evaluated on the datasets DCASE-2019~\citep{mesaros2018multi} and DCASE-2018~\citep{fonseca2018general}, obtaining an accuracy of up to 76.4\%, and 95.4\%, respectively, in both cases lower than the state of the art. Whereas for music genre classification, PANN achieved an accuracy of 91.5\% on the dataset GTZAN~\citep{DBLP:journals/taslp/TzanetakisC02}, lower than the state of the art. In all cases, the accuracy reported is higher than or close to the state-of-the-art results.

Approaches based on embedding generation have demonstrated competitive performance in multiple audio processing tasks using transfer learning. However, all of them are based on a single representation of the original audio. Therefore, enrichment of the deep representations by another representation could improve the performance of such models.

Research on audio processing has been focused significantly on the use of a single representation of the audio. Among the most common representations are the use of spectrograms~\citep{DBLP:conf/tencon/MulimaniK18, DBLP:journals/mta/ZengMPY19, DBLP:conf/ami/SarthakSM19}, and MFCCs~\citep{DBLP:conf/slsp/LeeJ18, DBLP:conf/icapr/AhmadTNP15}, which can be competitive depending on the task and the dataset used, and, in general, both can obtain similar results~\citep{DBLP:conf/slt/MeghananiSR21}. 

Different representations and features extracted from an audio can be used at the same time to feed a model. One example is FuzzyGCP~\citep{garain2021fuzzygcp}, which is a model fed by eight types of representations generated from the original audios and which are joined into a single two-dimensional image. FuzzyGCP was evaluated for language identification on the datasets IIIT Hyderabad~\citep{DBLP:conf/interspeech/PrahalladKKSB12}, IIT Madras~\citep{baby2016resources}, VoxForge, and MaSS~\citep{DBLP:conf/lrec/BoitoHGFB20}, obtaining accuracies of 95\%, 81.5\%, 68\%, and 98.7\%, respectively. These results exceeded the ones obtained by other state-of-the-art approaches, such as PPRLM~\citep{DBLP:conf/icassp/ZissmanS94}, i-vector~\citep{DBLP:conf/asru/SnyderGP15} and x-vector~\citep{DBLP:conf/icassp/SnyderGSPK18}.

The combination of representations makes possible to extract complementary information from the original audios, in a format easily processed by a machine learning model. This allows these models to achieve better results than those obtained by being fed by a single representation.

Another model based on the combination of representations is the model proposed by~\cite{DBLP:journals/corr/abs-2002-09607}, in which they combined three types of audio representations, which fed two models, one trained for acoustic scene classification and the other for general audio tagging. They use the DCASE 2018 Challenge dataset\footnote{http://dcase.community/challenge2018}, achieving a mAP@3 of 93.3\% in the acoustic scene classification task and an accuracy of 72.48\% in the acoustic scene classification task, outperforming the results of other state-of-the-art methods based on a single representation.

\cite{DBLP:journals/taslp/ZhuXKWP20} proposed a novel architecture fed by three types of representations, these representations fed two consecutive NN. One network is responsible for identifying and filtering erroneously labeled training data so that they do not affect the training of the other network, thus avoiding data errors that may adversely affect the performance of the model. They tested their architecture in audio tagging with the FSDKaggle2018\footnote{https://zenodo.org/record/2552860} and FSDKaggle2019\footnote{https://zenodo.org/record/3612637} datasets, each one evaluated with a different metric, achieving a mAP@3 of 95.59\%, and a label-weighted label-ranking average precision (lwlrap) of 0.7195 respectively, being, in both cases, competitive with the state-of-the-art methods.

We propose the MeWEHV architecture that enriches the embeddings generated by a pre-trained wave encoder model by combining it with embeddings extracted from MFCC representations through specialized neural layers in the architecture. Using the combination of both types of embeddings we are able to surpass the state-of-the-art results in multiple speech processing tasks, taking advantage of the benefits of embedding generation models and combination of representations.

\section{YouSpeakers204 dataset} \label{section:youspeaker}

We introduce a new dataset for speaker identification, called YouTube Speakers $204$ (YouSpeakers204), which contains $19607$ audio clips of $204$ speakers with $6$ different accents extracted from YouTube videos. We selected YouTube channels in which the information of the country of origin and gender of the speaker was stated and looked for native English speakers with a wide range of ages, ethnicities, and professions. Their respective YouTube channels contain different topics. All speakers are native English speakers, grouped by region of origin. The $6$ labeled regions are the United States, Canada, Scotland, England, Ireland, and Australia. The dataset is gender-balanced, with $50$\% male speakers and $50$\% female speakers. The audios included in this dataset come from varied recording environments, including indoor studios, outdoor recordings, professional recordings, and recordings with background noise.

The presented dataset is intended to facilitate research in the field of automatic speaker classification, and can also be used in related studies combining speaker identification with accent identification. 

YouSpeakers204 contains data recorded in noisy environments under a wide variety of real conditions, such as the recording microphone used, background noise, recording environment, audio volume, speaker gender, and accent, which makes the dataset a challenge to test the robustness of new machine learning models.

\begin{figure}[thpb]
  \centering
  {\includegraphics[scale=0.35]{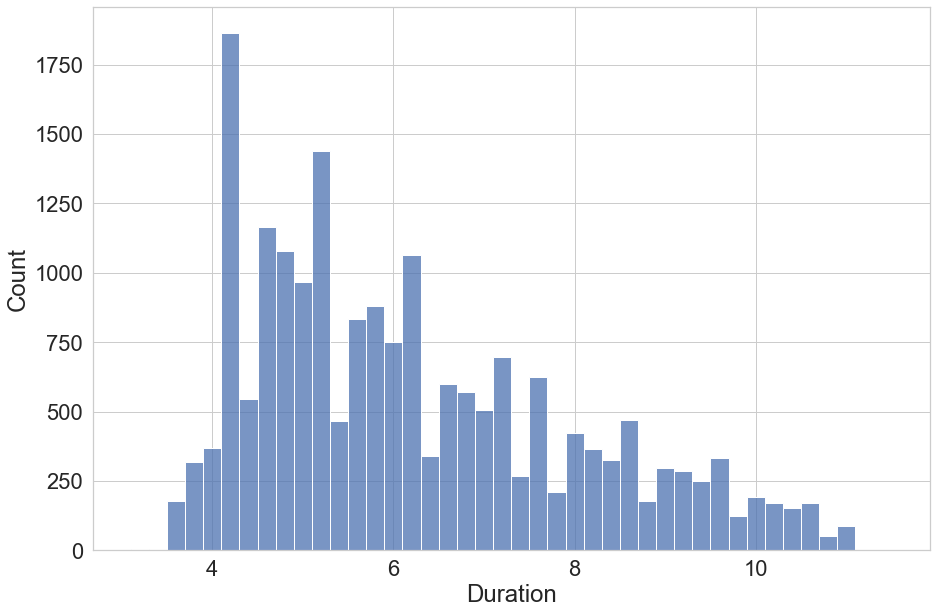}
  }
  \caption{Distribution of clip duration in seconds in the YouSpeakers204 dataset.  The audios are between 3.5 seconds and 12 seconds in length.}
  \label{figure:stats_lenght_1}
\end{figure}

The general statistics of the dataset are presented in Table~\ref{table:stats_dataset}, while Fig.~\ref{figure:stats_lenght_1} includes the length distribution of the audio clips.

\begin{table}[h]
\caption{YouSpeakers204 dataset statistics}
\label{table:stats_dataset}
\begin{center}
\begin{tabular}{rl}
\hline
\# of speakers & 204 \\ 
\# of male speakers & 102 \\ 
\# of female speakers & 102 \\ 
\# of accents & 6 \\ 
\# of videos & 1055 \\ 
\# of minutes & 2026 \\ 
\# of audio clips & 19607 \\ 
Avg. \# of videos per speaker & 5.17 \\ 
Avg. \# of clips per speaker & 96.11 \\ 
Avg. length of clips & 6.19 seconds \\ \hline
\end{tabular}
\end{center}
\end{table}

For the creation of YouSpeakers204, we defined a procedure, which consists of the following stages: listing candidate speakers, selecting and downloading videos, and audio processing.

\subsection{Listing candidate speakers}
The list of speakers was extracted from the Socialblade\footnote{https://socialblade.com} database by selecting the most famous YouTubers for each region among the $6$ accent classes. In the case of Scotland and England, it was necessary to perform a manual search within YouTube to find people who, in their public information, claim to be from those regions, because Socialblade divides YouTubers by countries and not by regions. Subsequently, a verification of the place of birth of each speaker was carried out, by a search in Wikipedia\footnote{https://www.wikipedia.org}. The collected list contains a total of $204$ speakers, $34$ per accent of which $17$ are men and $17$ are women ($(17+17)\times6=204$). All speakers have been assigned a unique identifier (id) and their real identity is not provided in the dataset.

\subsection{Selecting and downloading videos}

To create a diverse dataset, $19607$ clips were extracted from a large number of videos ($1055$ videos in total). In this way, models created using YouSpeakers204 dataset can be robust to the different environments and contexts in which the audios had been recorded. Each speaker has an average of $5.17$ videos from which their clips were extracted, an average of $96.11$ clips, and each clip has an average length of $6.19$ seconds.

After selecting the videos, the entire video was downloaded and the audio clips were extracted.

\subsection{Audio processing}

\begin{figure}[thpb]
  \centering
  {\includegraphics[width=\textwidth]{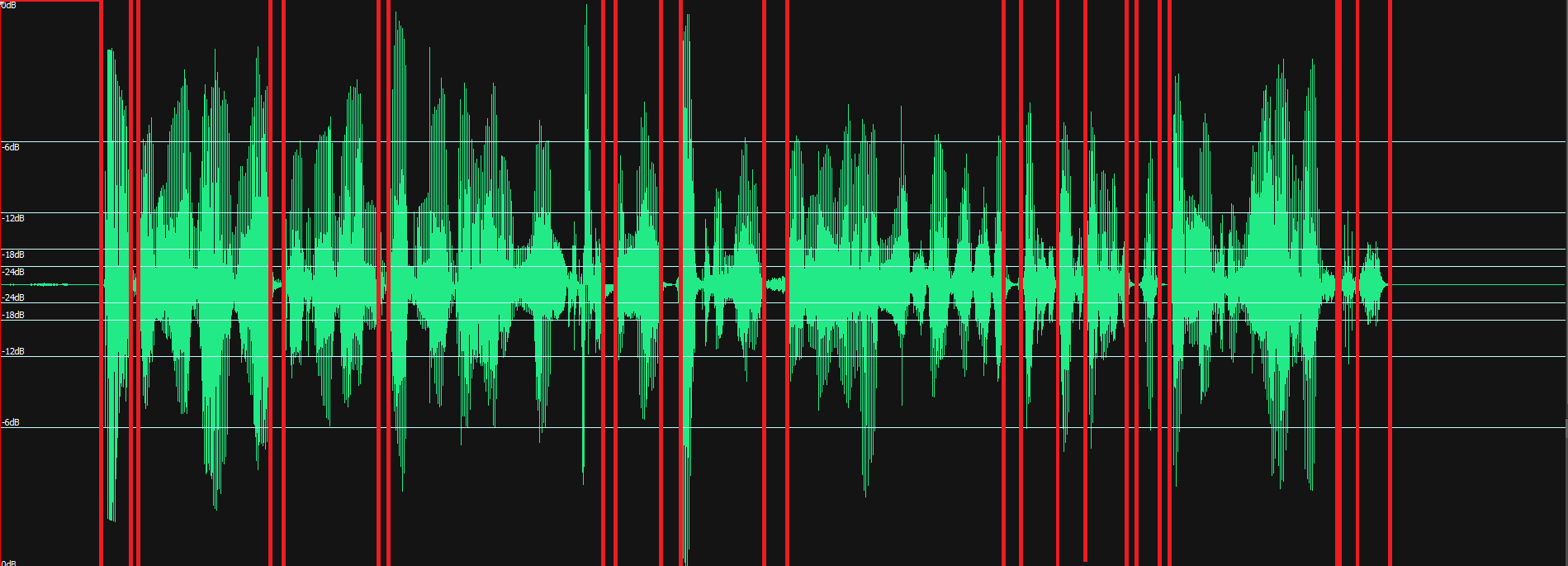}
  }
  \caption{Example of an audio cut-off in ``silence" regions, i.e. regions where the decibels are below a certain threshold. The audio waveform is shown in green and the cut-off regions are in red.}
  \label{figure:audio_cut_silence}
\end{figure}

The complete audios were processed manually, by a team of taggers, separating the original audios into segments of short duration and storing the resulting audios together with their respective information. The process consisted of defining a decibel split threshold for each audio, all generated sections with a decibel level below the threshold are considered silences (see Fig.~\ref{figure:audio_cut_silence}). All segments are obtained by extracting the sections that are between every two contiguous silences. Due to the particularities of each audio, it is not possible to define a unique threshold that allows to label the silences of all the audios in a correct way. Therefore, applying a visual analysis of the audio waves the threshold of each file was defined manually.

The segmentation of the audios generates multiple clips of variable size, of which only the clips with a duration between $3.5$ and $12$ seconds are kept. A manual check of the content of each clip is then performed to discard all clips containing voices other than the target YouTuber, and clips containing no voice.

Finally, the resulting clips are renamed and stored, the names of the files contain an anonymized speaker id, anonymized id of the video from which each clip originates, the gender of the speaker, and the region of the speaker, which represents the accent.

\section{Architecture: Mel and Wave Embeddings for Human Voice Tasks} \label{section:architecture}


%
%

\begin{figure}[thpb]
  \centering
  {\includegraphics[width=\textwidth]{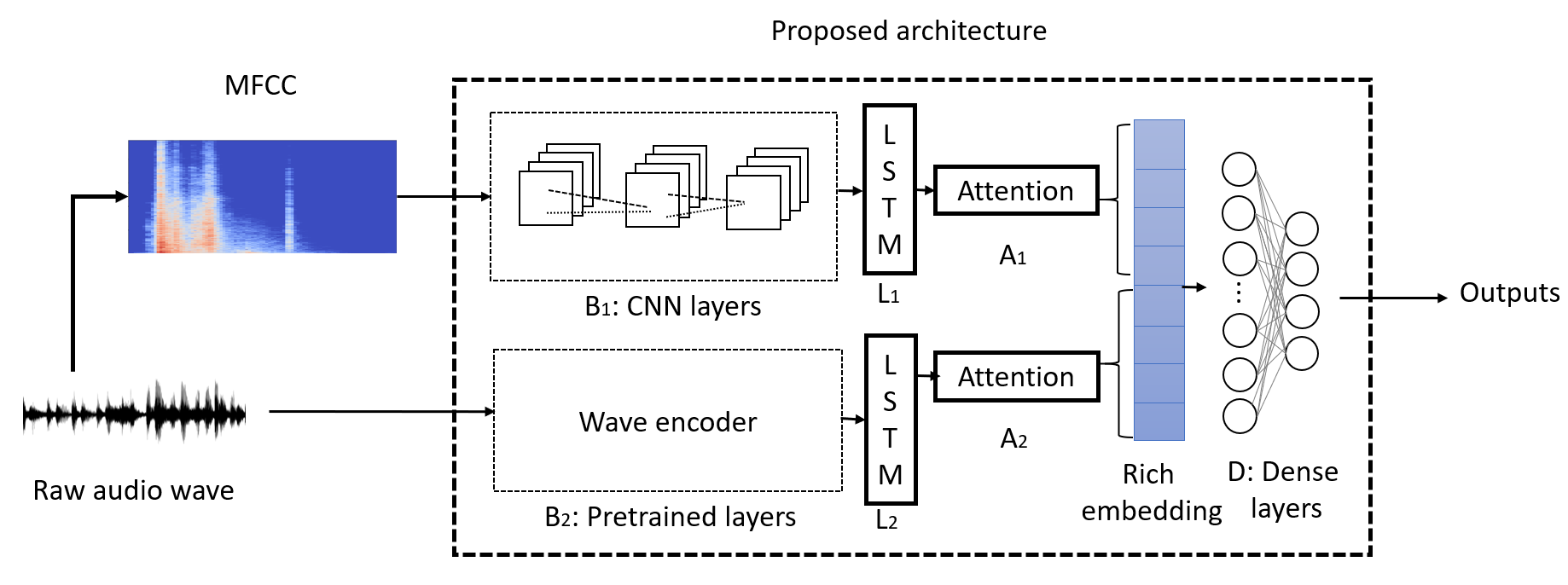}
  }
  \caption{The architecture of the MeWEHV model proposed in this paper. The model is fed by two inputs, on the one hand, the raw audio waveform that feeds the wave encoder branch, which contains the encoder layers of a pre-trained embedding generation model from raw audios.  On the other hand, the MFCC coefficients extracted from the same raw audio feed the MFCC branch which contains a set of convolutional layers. Both branches of the model pass through an LSTM layer and a Soft Attention layer independently. These layers are in charge of modelling the temporal features and generating, each one, a unique embedding from each input audio. Both embeddings are concatenated and subsequently feed a series of dense layers that are responsible for the classification of the model.}
  \label{figure:pipeline_model_full}
\end{figure}

Figure~\ref{figure:pipeline_model_full} depicts a summary of the Mel and Wave Embeddings for Human Voice Tasks (MeWEHV) architecture.

The MeWEHV architecture is fed by two inputs, on the one hand, the audio signal is treated as a one-dimensional vector, and on the other hand, the MFCCs~\citep{tiwari2010mfcc} are extracted from the same audio signal.

\paragraph*{Wave encoder branch} The two inputs of the MeWEHV model are processed independently by two branches. On the one hand, the raw audio waveform feeds the encoder layers of a $B_{2}$ wave encoder model. The $B_{2}$ wave encoder is a block of the architecture that is composed of a pre-trained embedding generation model. In this paper, multiple models were tested as the wave encoder block.


The models used as wave encoders are XLSR-Wav2Vec2, HuBERT in its base and large versions, and WavLM in its base and large versions. In future research, these models could be replaced by others.

The layers of the wave encoder models can be grouped into two categories, according to their functionality. On the one hand, the encoder layers take as input a raw audio, divide it into $25$ ms sections $S={S_{1}, S_{2}, ... , S_{N}}$ and generate a set of embeddings $E={E_{1}, E_{2}, ... , E_{N}}$, one per section. Each embedding $E_{i}$, where $1\leq i \leq N$ represent features of a section $S_{i}$ that were relevant during the training process for its target task. On the other hand, the decoder layers take the embeddings and process them to generate the speech transcription. 

The embeddings generated by the wave encoder represent a position in latent space, where audios with similar characteristics are represented spatially close together, and audios with different characteristics are represented far apart.

The first branch of our model will generate multiple embeddings $E_{i}$, one for each audio section $S_{i}$, which will summarize the features that the wave encoder considers relevant. We perform transfer learning by feeding the generated embeddings to new layers connected to the encoder outputs.

We connected the embeddings of the generated sections to an LSTM layer, $L_{2}$, and a Soft Attention layer, $A_{2}$ that will be able to model their temporal information and generate a single embedding for the complete audio signal. The blocks that compose this branch are $B_{2} + L_{2} + A_{2}$.

\paragraph*{MFCC branch} The second branch of the proposed model is fed by the information contained in the MFCCs of the original audios and generates a new embedding. In this way we enrich the embeddings generated by the first branch of the model.

The MFCCs are coefficients based on the human audible frequency range, represented by the Mel scale, which is a linear scale below $1000$ Hz and logarithmic above $1$ kHz~\citep{tiwari2010mfcc}. The Mel scale is a scale that relates the perceived frequency of a tone to the actual measured frequency. It modifies a frequency to approximate what the human ear can hear, and is often used to extract features of an audio signal that are relevant to identifying its content, making it useful in tasks such as speech representation.

The MFCCs reduce the relevance of information that may have a minor contribution to speech processing tasks, which may add noise to the model and reduce its accuracy. It allows us to analyze the most important information of an audio and complement the information that the wave encoder models may have missed during their analysis.

We use the $B_{1}$ block, composed of three concatenated layer sets to process the MFCCs, each set consists of a 1D convolutional layer, a batch normalization layer, and a ReLU activation function. At the same time, the output of the last block is connected to an LSTM layer, $L_{1}$, and a Soft Attention layer, $A_{1}$, in the same way as the first branch of the model. The output of this layer block is a new embedding.The blocks that compose this branch are $B_{1} + L_{1} + A_{1}$.

\paragraph*{Rich embeddings} The embedding generated by the first branch and the embedding generated by the second branch are concatenated and generate a new embedding which size is the sum of the embedding sizes of both branches. All trainable parameters in the previous layers are trained using center loss, specially designed for embedding generation~\citep{DBLP:conf/eccv/WenZL016}. Thus, the result of the concatenation of both embeddings is a rich embedding in a new latent space.

Finally, the rich embedding feeds a block $D$ of two fully connected neuron layers, with intermediate ReLU activation and a dropout function, which is responsible for generating the output of the model that classifies the input according to the assigned task. The block $D$ is trained using negative log likelihood loss~\citep{DBLP:journals/corr/abs-1804-10690}. 

The proposed MeWEHV architecture complements the information of the pre-trained wave encoder model with the information extracted using the MFCCs to generate a more powerful and flexible model.

\section{Experiments} \label{section:experiments}

\subsection{Datasets}

The datasets used allow us to evaluate the performance of our model in the tasks of language identification, accent identification, and speaker identification. These datasets are diverse in terms of the number of speakers, nationalities, gender, and environments in which they were recorded, which allows us to evaluate the correct functionality of a MeWEHV model with complex data.

The lists of audios used in the training, validation, and test partitions are publicly available\footnote{Partitions of the datasets used are publicly available at: https://bit.ly/3ydSEAt}, so that future research can make a fair comparison of results.

\subsubsection{VoxForge}

VoxForge is a dataset composed of the voices and transcriptions of a large number of speakers, originally intended for speech-to-text conversion. It comprehends a large number of languages, which makes it also useful for language identification. 

We use this dataset to compare the proposed architecture in language identification. The used subset is based on the FuzzyCGP paper presented by \cite{garain2021fuzzygcp}. 

This subset contains $5$ languages: French, German, Italian, Portuguese, and Spanish. All the speakers of each language were divided into a proportion of $70$\% for training, $10$\% for test, and $20$\% for validation. Subsequently, considering only the audios with the selected speakers, $1400$ audio clips were randomly chosen for the training set, $200$ for the test set, and $400$ for the validation set, per language. This results in a training, test and validation sets with $7000$, $1000$ and $2000$ audios, respectively.

The generated partitions have no speaker contamination, i.e., the speakers present in one set are not present in the other sets, which assures that the model learns to recognize the languages and not the voices of the speakers.

\subsubsection{Common Language}

The Common Language dataset~\citep{ganesh_sinisetty_2021_5036977} has a set of audios selected from the Common Voice dataset~\citep{DBLP:conf/lrec/ArdilaBDKMHMSTW20}, which contains audios provided by volunteers. Common Language contains 45 languages, 272360 audios, and 13808 speakers.

The dataset was used for language identification and we used the training, validation, and test partitions provided by the authors: $177552$, $47104$ and $47704$ audios for training, validation and test sets, respectively.

\subsubsection{Latin American Spanish Corpora}

Latin American Spanish Corpora~\citep{guevara-2020-crowdsourcing} was originally proposed for the speech-to-text conversion task. However, thanks to being a highly balanced dataset, both by gender and by accents, it can be used in accent identification. In this paper, to the best of our knowledge, this is the first time that this dataset is being used for for accent identification.

The dataset comprises $37.79$ hours of $6$ Latin American accents: Argentinian, Chilean, Colombian, Peruvian, Puerto Rican and Venezuelan. We divided the speakers of the dataset into training ($70$\%), validation ($15$\%) and test sets ($15$\%), and use all the audios of each speaker.

\subsubsection{Common Voice}

We used the Common Voice dataset for the task of accent identification. We worked with an English subset containing audios of five accents: American, British, Indian, Canadian, and Australian. 

We used a subset with 10000 audios per accent, which were divided into training (70\%), validation (15\%) and test sets (15\%), resulting in a training set with 35000 audios, a validation set with 7500 audios, and a test set with 7500 audios.

\subsubsection{VoxCeleb1}

For the speaker identification task, another of the datasets we chose is VoxCeleb1~\citep{ahamad-anand-bhargava:2020:LREC}, which is composed of $153516$ audios samples from $1251$ different speakers. The audios of the dataset were extracted from public videos of celebrities on YouTube.

For the VoxCeleb1 dataset, the same partition proposed by the creators~\footnote{$https://www.robots.ox.ac.uk/\sim vgg/data/voxceleb/meta/iden\_split.txt$} was used, which contains $138361$ audios in the training set, $6904$ audios in the validation set, and $8251$ audios in the test set.

\subsubsection{YouSpeakers204}

The YouSpeakers204 dataset is one of the contributions of this paper and its information can be found in Section~\ref{section:youspeaker}.

The entire YouSpeakers204 dataset was used for speaker identification.
The audios of each speaker were divided into a proportion of $70$\% for training, $15$\% for validation, and $15$\% for test, and all the audios available in the dataset were used. This resulted in a training set of $13728$ audios, $2942$ audios in the validation set, and $2942$ audios in the test set.

\subsection{Experimental setup}

For the experimentation, in all the datasets and tasks, audios with a sample rate of $16000$ samples per second were used, converted into $8$-second clips. Those with a shorter duration were repeated as many times as necessary until reaching $8$ seconds, and those with longer duration were trimmed and only the first $8$ seconds were worked on. 

In the creation of the MFCCs, $128$ MFCC coefficients were defined as a parameter to be used, which we consider it provides the MFCC with a high level of spectral detail, which allows the models to perform tasks requiring such detail.

The specific parameters of the MeWEHV model used in our experimentation can be seen in Table~\ref{table:proposed_model_architecture}, which were empirically selected.

\begin{table*}[thpb]
\caption{Details of the MeWEHV model used in the experimentation, assuming a task with six possible output classes. \textit{(T)} represents that the input of a given layer has been transposed. With each extra output class, the number of parameters increases by 256, being the number of connections that the new output neuron would have with the penultimate layer. The number of wave encoder parameters depends on the model chosen for that block, and the size of the generated embeddings of 1024 was assumed as the wave encoder output.}
\label{table:proposed_model_architecture}
\begin{center}
\resizebox{\textwidth}{!}{%
\begin{tabular}{|llcccc|}
\hline
\multicolumn{1}{|c|}{\textbf{\begin{tabular}[c]{@{}c@{}}Layer blocks \\ (index. id: type: depth)\end{tabular}}} &
  \multicolumn{1}{c|}{\textbf{Layers}} &
  \multicolumn{1}{c|}{\textbf{\begin{tabular}[c]{@{}c@{}}Input \\ shape\end{tabular}}} &
  \multicolumn{1}{c|}{\textbf{\begin{tabular}[c]{@{}c@{}}Output \\ shape\end{tabular}}} &
  \multicolumn{1}{c|}{\textbf{Param \#}} &
  \textbf{\begin{tabular}[c]{@{}c@{}}Kernel \\ shape\end{tabular}} \\ \hline
\multicolumn{1}{|l|}{\multirow{9}{*}{1. $B_{1}$: CNN: 1}} &
  \multicolumn{1}{l|}{Conv1d} &
  \multicolumn{1}{c|}{{[}128, 641{]}} &
  \multicolumn{1}{c|}{{[}128, 319{]}} &
  \multicolumn{1}{c|}{82,048} &
  {[}128, 128, 5{]} \\ \cline{2-6} 
\multicolumn{1}{|l|}{} &
  \multicolumn{1}{l|}{BatchNorm1d} &
  \multicolumn{1}{c|}{{[}128, 319{]}} &
  \multicolumn{1}{c|}{{[}128, 319{]}} &
  \multicolumn{1}{c|}{256} &
  {[}128{]} \\ \cline{2-6} 
\multicolumn{1}{|l|}{} &
  \multicolumn{1}{l|}{ReLU} &
  \multicolumn{1}{c|}{{[}128, 319{]}} &
  \multicolumn{1}{c|}{{[}128, 319{]}} &
  \multicolumn{1}{c|}{} &
   \\ \cline{2-6} 
\multicolumn{1}{|l|}{} &
  \multicolumn{1}{l|}{Conv1d} &
  \multicolumn{1}{c|}{{[}128, 319{]}} &
  \multicolumn{1}{c|}{{[}128, 316{]}} &
  \multicolumn{1}{c|}{65,664} &
  {[}128, 128, 4{]} \\ \cline{2-6} 
\multicolumn{1}{|l|}{} &
  \multicolumn{1}{l|}{BatchNorm1d} &
  \multicolumn{1}{c|}{{[}128, 316{]}} &
  \multicolumn{1}{c|}{{[}128, 316{]}} &
  \multicolumn{1}{c|}{256} &
  {[}128{]} \\ \cline{2-6} 
\multicolumn{1}{|l|}{} &
  \multicolumn{1}{l|}{ReLU} &
  \multicolumn{1}{c|}{{[}128, 316{]}} &
  \multicolumn{1}{c|}{{[}128, 316{]}} &
  \multicolumn{1}{c|}{} &
   \\ \cline{2-6} 
\multicolumn{1}{|l|}{} &
  \multicolumn{1}{l|}{Conv1d} &
  \multicolumn{1}{c|}{{[}128, 316{]}} &
  \multicolumn{1}{c|}{{[}128, 313{]}} &
  \multicolumn{1}{c|}{65,664} &
  {[}128, 128, 4{]} \\ \cline{2-6} 
\multicolumn{1}{|l|}{} &
  \multicolumn{1}{l|}{BatchNorm1d} &
  \multicolumn{1}{c|}{{[}128, 313{]}} &
  \multicolumn{1}{c|}{{[}128, 313{]}} &
  \multicolumn{1}{c|}{256} &
  {[}128{]} \\ \cline{2-6} 
\multicolumn{1}{|l|}{} &
  \multicolumn{1}{l|}{ReLU} &
  \multicolumn{1}{c|}{{[}128, 313{]}} &
  \multicolumn{1}{c|}{{[}128, 313{]}} &
  \multicolumn{1}{c|}{} &
   \\ \hline
\multicolumn{2}{|l|}{2. $L_{1}$: LSTM: 2} &
  \multicolumn{1}{c|}{{[}313, 128{]} (T)} &
  \multicolumn{1}{c|}{{[}313, 128{]}} &
  \multicolumn{1}{c|}{132,096} &
  {[}128, 128{]} \\ \hline
\multicolumn{2}{|l|}{3. $A_{1}$: SoftAttention: 3} &
  \multicolumn{1}{c|}{{[}313, 128{]}} &
  \multicolumn{1}{c|}{{[}128{]}} &
  \multicolumn{1}{c|}{16,512} &
  {[}128, 128{]} \\ \hline
\multicolumn{2}{|l|}{4. $B_{2}$: Wave encoder: 1} &
  \multicolumn{1}{c|}{{[}1, 128000{]}} &
  \multicolumn{1}{c|}{{[}399, 1024{]}} &
  \multicolumn{1}{c|}{variable} &
   \\ \hline
\multicolumn{2}{|l|}{5. $L_{2}$: LSTM: 2} &
  \multicolumn{1}{c|}{{[}399, 1024{]}} &
  \multicolumn{1}{c|}{{[}399, 128{]}} &
  \multicolumn{1}{c|}{590,848} &
  {[}1024, 128{]} \\ \hline
\multicolumn{2}{|l|}{6. $A_{2}$: SoftAttention: 3} &
  \multicolumn{1}{c|}{{[}399, 128{]}} &
  \multicolumn{1}{c|}{{[}128{]}} &
  \multicolumn{1}{c|}{16,512} &
  {[}128, 128{]} \\ \hline
\multicolumn{1}{|l|}{\multirow{5}{*}{7. $D$: Dense layers: 4}} &
  \multicolumn{1}{l|}{Linear} &
  \multicolumn{1}{c|}{{[}256{]}} &
  \multicolumn{1}{c|}{{[}256{]}} &
  \multicolumn{1}{c|}{65,536} &
  {[}256, 256{]} \\ \cline{2-6} 
\multicolumn{1}{|l|}{} &
  \multicolumn{1}{l|}{ReLU} &
  \multicolumn{1}{c|}{{[}256{]}} &
  \multicolumn{1}{c|}{{[}256{]}} &
  \multicolumn{1}{c|}{} &
   \\ \cline{2-6} 
\multicolumn{1}{|l|}{} &
  \multicolumn{1}{l|}{Dropout} &
  \multicolumn{1}{c|}{{[}256{]}} &
  \multicolumn{1}{c|}{{[}256{]}} &
  \multicolumn{1}{c|}{} &
   \\ \cline{2-6} 
\multicolumn{1}{|l|}{} &
  \multicolumn{1}{l|}{Linear} &
  \multicolumn{1}{c|}{{[}256{]}} &
  \multicolumn{1}{c|}{{[}6{]}} &
  \multicolumn{1}{c|}{1536} &
  {[}256, 6{]} \\ \cline{2-6} 
\multicolumn{1}{|l|}{} &
  \multicolumn{1}{l|}{LogSoftmax} &
  \multicolumn{1}{c|}{{[}6{]}} &
  \multicolumn{1}{c|}{{[}6{]}} &
  \multicolumn{1}{c|}{} &
   \\ \hline
\multicolumn{6}{|l|}{Non-trainable params: wave encoder params} \\ \hline
\multicolumn{6}{|l|}{Trainable params: 1,038,982} \\ \hline
\multicolumn{6}{|l|}{Total params: wave encoder params + 1,038,982} \\ \hline
\end{tabular}
}
\end{center}
\end{table*}

We established multiple baseline models, on which the MeWEHV architecture is applied, and took them as a reference to compare the performance of our proposal. These architectures are CNNMFCC (composed of a CNN fed by MFCCs and the classification layers of the second branch), and wave encoders such as XLSR-Wav2Vec2, WavLM base, WavLM large, HuBERT base and HuBERT large.

The wave encoder architectures are composed only of the pre-trained embedding generation model encoder and the classification layers of the first branch. It is made up of neuron blocks $B_{2} + L_{2} + A_{2} + D$, which can be seen in Table~\ref{table:proposed_model_architecture}. Wave encoder architectures use $128$ neurons instead of $256$ in block $D$ because only one branch is used, resulting in a model with $625,542$ trainable parameters.

While the CNNMFCC comprises neuron blocks $B_{1} + L_{1} + A_{1} + D$ because only one branch is used. CNNMFCC use $128$ neurons instead of $256$ in the layers block $D$, resulting in a model with $355,654$ trainable parameters.

\section{Results} \label{section:results}

The results of the experiments performed can be seen in Table~\ref{table:results}. 

\begin{table*}[thpb]
\caption{Results in terms of accuracy, obtained in three speech classification tasks, with six datasets. The models starting with the designation ``MeWEHV-X" refer to the models in which the proposed architecture was applied using the baseline model ``X" as a wave encoder. The acronym CL refers to the Common Language dataset, CV refers to the Common Voice dataset, LASC refers to the Latin American Spanish Corpora dataset, and YS204 refers to the YouSpeakers204 dataset. The best result of each pair of ``X" and ``MeWEHV-X" models is shown in italics, and the best overall results of each dataset are shown in bold.}
\label{table:results}
\resizebox{\textwidth}{!}{%
\begin{tabular}{@{}llcccccc@{}}
\toprule
\multicolumn{1}{c}{\multirow{2}{*}{Model}} & \multirow{2}{*}{\# Params.} & \multicolumn{2}{c}{Language}       & \multicolumn{2}{c}{Accent}        & \multicolumn{2}{c}{Speaker}           \\ \cmidrule(l){3-8} 
\multicolumn{1}{c}{}                       &                          & \multicolumn{1}{l}{VoxForge} & \multicolumn{1}{l}{CL} & \multicolumn{1}{l}{LASC} & \multicolumn{1}{l}{CV} & \multicolumn{1}{l}{VoxCeleb1} & \multicolumn{1}{l}{YS204} \\
\toprule
CNNMFCC                                    & 0.36M                    & 67.15\%                  &            18.88\%            & 84.63\%                 &         33.77\%               &          41.79\%                & 63.82\%                   \\
\toprule
XLSR-Wav2Vec2                              & 318.00M                  & 80.70\%                  &        37.06\%                & 85.39\%                 &         33.39\%               &           49.18\%               & 83,27\%                   \\
\textbf{MeWEHV-XLSR-Wav2Vec2}                & 318.42M                  & \textit{92.70\%}                  &  \textit{38.31\%}              & \textbf{87,62\%}        &         \textit{33.69\%}               &              \textit{64.40\%}            & \textit{87,96\%}                   \\
\toprule
HuBERT-base                                & 95.30M                   & 87.09\%                  &       39.89\%                 & 77.43\%                 &         32.09\%               &         51.66\%                 & 59.03\%                   \\
\textbf{MeWEHV-HuBERT-base}                  & 95.72M                   & \textit{94.19\%}                  &         \textit{44.87\%}
               & \textit{82.68\%}                 &            \textit{38.63\%}            &          \textit{64.86\%}                & \textit{87.35\%}                   \\
\toprule
HuBERT-large                               & 317.23M                  & 85.90\%                  &        46.55\%                & 67.89\%                 &         34.84\%               &          40.87\%                & 46.65\%                   \\
\textbf{MeWEHV-HuBERT-large}                 & 317.65M                  & \textit{94.49\%}         &        \textit{49.45\%}                &  \textit{81.59\%}        &        \textit{35.73\%}                &      \textit{64.86\%}
                    &  \textit{87.83\%}          \\
\toprule
WavLM-base                                 & 95.32M                   & 92.50\%                  &         57.33\%               & 83.29\%                 &             35.35\%           &           47.55\%               & 61.27\%                   \\
\textbf{MeWEHV-WavLM-base}                   & 95.74M                   & \textit{97.30\%}                  &          \textit{62.51\%}              & \textit{83.49\%}                 &          \textit{36.40\%}              &        \textit{67.09\%}                  & \textit{87.49\%}                   \\
\toprule
WavLM-large                                & 317.24M                  & 96.89\%                  &          63.36\%              & 80.20\%                 &        38.60\%               &            61.49\%              & 67.53\%                   \\
\textbf{MeWEHV-WavLM-large}                  & 317.66M         & \textbf{97.60\%}         & \textbf{72.53\%}              & \textit{83.63\%}                 &          \textbf{42.55\%}              &          \textbf{70.62\%}                & \textbf{89.22\%}          \\ \bottomrule
\end{tabular}
}
\end{table*}

The number of parameters mentioned in Table~\ref{table:results} for the different versions of XLSR-Wav2Vec2, HuBERT and WavLM include the 0.62M trainable parameters of the LSTM, Soft Attention, and classification layers added, which have as input the embeddings generated by each of the mentioned models.

As it can be seen, the implemented MeWEHV models improve the results with respect to all the embedding generation models used as wave encoders and on which our proposal was implemented. It is worth noting that the MeWEHV models have only 0.68M more parameters than their corresponding baseline models, which represents, in the case of the XLSR-Wav2Vec2 model, the baseline model with the highest number of parameters, an increase of only $0.21\%$ of parameters. 

On the VoxForge and Common Language datasets, the best language identification model is MeWEHV, using WavLM-large as wave encoder, achieving accuracies of $97.06\%$ and $72.53\%$, respectively, which represents an improvement of $0.73\%$ and $14.47\%$ with respect to WavLM large, with the improvement achieved with Common Language being the largest among the models tested in this dataset with respect to its baseline model. The largest improvement with VoxForge was achieved with the XLSR-Wav2Vec2-based MeWEHV model with respect to the XLSR-Wav2Vec2 model and represents an increase of $14.86\%$.  

In addition to the mentioned results, we can add as a baseline the result achieved by the FuzzyGCP model in language identification with the VoxForge dataset, obtaining an accuracy of $68\%$, being particularly relevant since the FuzzyGCP model is based on another approach for input combination with multiple audio representations. Our best MeWEHV model has an improvement of up to $43.53\%$ over the result achieved by FuzzyGCP, with a VoxForge subset inspired by the one used in the FuzzyGCP paper.

On the YouSpeakers204 and VoxCeleb1 datasets, the best speaker identification model is again MeWEHV using WavLM-large, yielding accuracies of $89.22\%$ and  $70.62\%$, respectively. The largest improvements were achieved with the HuBERT-large-based MeWEHV model with respect to the HuBERT-large model and represent an increase of $88.27\%$ and $58.69\%$, respectively.  

In the case of the accent identification task, the best model on the Common Voice dataset is MeWEHV based on WavLM-large, and on the LASC dataset is the MeWEHV model based on XLSR-Wav2Vec2, which achieve accuracies of $42.55\%$ and $81.59\%$ respectively. On Common Voice the highest improvement is $20.38\%$ and was obtained with the MeWEHV-HuBERT-base model with respect to HuBERT-base and on the LASC dataset the highest improvement is $20.18\%$ and was achieved with MeWEHV-HuBERT-large with respect to HuBERT-large.

The MeWEHV model based on WavLM-large outperformed the MeWEHV model based on XLSR-Wav2Vec2 in language identification.
However, in the case of accent classification in Spanish with the LASC dataset, the capabilities of MeWEHV based on XLSR-Wav2Vec2 can be appreciated, outperforming the other models.

The experiments showed a large increase of accuracy for speaker identification when MeWEHV is used, especially in the models based on the different versions of HuBERT and WavLM. The MeWEHV models based on these wave encoders can be considered a fusion between these wave encoders and the CNNMFCC model, therefore, we can notice that the fusion of both approaches significantly exceed the performance of each approach separately and that the resulting model takes advantage of the modeling capabilities of both.

\section{Discussion} \label{section:discussion}

In the proposed MeWEHV architecture, we used two modules that work together to generate rich embeddings. On the one hand, a module for extracting features from raw audios using multiple embedding generation models. On the other hand, a module obtaiing more features using a series of convolutional layers fed by the MFCCs of the original audios. The joint work of both modules proved to achieve better results than those obtained by the two modules separately. To validate our approach, we experimented with six datasets, used for three different tasks, two datasets per task.

Our studies show that optimal results can be obtained after combining both types of inputs in a single architecture and generating rich embeddings. This relates to the findings presented by~\cite{garain2021fuzzygcp} on FuzzyGCP, where although an approach based on the generation of embeddings was not used, it was shown that combining different audio representations can improve the results obtained with each of these representations individually. In the language identification task with the VoxForge dataset, our approach, based on embeddings generation, proved to be able to achieve better results.

We compared our proposal with five state-of-the-art models and found that the MeWEHV version of each model was able to achieve superior accuracy on all the datasets used. In addition to this, we found that one of the advantages of the proposed model is that only a small number of new parameters are required to be learned to significantly increase the performance of the baseline models.

The MFCC branch of the MeWEHV architecture contains only three convolutional layers. Given that in a CNN, the first layers are in charge of modeling the low-level features, such as straight lines, edges, and corners, and the later layers are in charge of modeling the high-level features, we can conclude, based on the results obtained, that the addition of low-level features from the MFCC works well complementing the information extracted by the wave encoders.

We can also note that the only model that managed to outperform the MeWEHV-WavLM-large model was the MeWEHV-XLSR-Wav2Vec2 model in the specific case of accent identification with the LASC dataset. This may be because XLSR-Wav2Vec2 was trained to be able to model multiple languages, so it should be able to identify different types of pronunciations and the use of different phonemes, in addition to those used in English, while the other models are only specialized in this language. Also noteworthy is the robustness of the XLSR-Wav2Vec2 model in speaker identification with the YouSpeakers204 dataset, being notably that this, baseline without MeWEHV, has the highest accuracy among all the baselines evaluated, but always lower than our proposal.

\section{Conclusions and future work} \label{section:conclutions}

In this work we have proposed MeWEHV, a machine learning model architecture that enriches the embeddings, generated by a pre-trained wave encoder, using features extracted from MFCC representations. MeWEHV was tested on the language identification task with the VoxForge, and Common Language datasets, achieving an accuracy of up to $97.60\%$ and $72.53\%$, respectively, superior to other state-of-the-art approaches. It should be noted that the MeWEHV architecture only requires $1.04M$ additional parameters in addition to the wave encoder parameters, representing only $0.33\%$ to $1.09\%$ additional parameters.

Furthermore, the model was tested in the identification of accents with the Latin American Spanish Corpora, achieving an accuracy of up to $87.62\%$. This is the first result reported with this dataset in this specific task, which will allow future research to have a reference result to compare with. Moreover, it was tested on the Common Voice dataset achieving an accuracy of up to $42.55\%$. In both datasets the MeWEHV models achieved the highest results.

We proposed YouSpeakers204, a new speaker identification dataset, highly balanced by accent and speaker gender,  in which MeWEHV obtained $89.22\%$, which is the highest accuracy. Together with the dataset, we proposed training, test, and validation sets, which can be used by other researchers for a fair comparison. In speaker identification, we also tested the VoxCeleb1 dataset, obtaining the best results with MeWEHV, with an accuracy of up to $70.62\%$.

In all experiments, the results of MeWEHV models were compared with the CNNMFCC model which is an MFCC-fed CNN model, and with five state-of-the-art models, i.e. WavLM base, WavLM large, XLSR-Wav2Vec2, HuBERT base and HuBERT large, outperforming, in all cases, their results. In this way, we demonstrated that our approach is superior to all baselines, in multiple speech classification tasks.

Thus, this work allows the use of a machine learning architecture that requires training with a relatively low additional computational cost and consistently achieves superior results than the baselines. Our architecture provides a general framework that can be used with other pre-trained models as wave encoders. 

In our proposal, we present an approach that combines wave encoders with convolutional layers fed by MFCC but for future research, we will analyze the performance achieved with other imposed representations.

\section*{Acknowledgements}
This research has been funded with support from the European Union’s Horizon 2020 Research and Innovation Framework Programme, H2020 SU-FCT-2019 under the GRACE project with Grant Agreement 883341. This publication reflects the views only of the authors, and the European Union’s Horizon 2020 Research and Innovation Framework Programme, H2020 SU-FCT-2019  cannot be held responsible for any use which may be made of the information contained therein.

This work was also supported by the framework agreement between the Universidad de Leon and INCIBE (Spanish National Cybersecurity Institute) under Addendum 01.
\bibliography{biblio}

\end{document}